\documentclass[conference]{IEEEtran}

\usepackage{amsmath,amssymb,amsfonts}
\usepackage{graphicx}
\usepackage{booktabs}
\usepackage{multirow}
\usepackage{array}
\usepackage{enumitem}
\usepackage{caption}
\usepackage{siunitx}
\usepackage{threeparttable}
\usepackage{cite}
\usepackage[hidelinks,bookmarks=false]{hyperref}
\begin{document}

\title{Shortcut in Generalization: Rethinking Evaluation Protocol for Encrypted Traffic Classification}

\author{
\IEEEauthorblockN{Qiming Yu\textsuperscript{1}, Chungang Lin\textsuperscript{2,3}, Zhengyi Ma\textsuperscript{4}, Zijia Song\textsuperscript{1}, Yuliang Lu\textsuperscript{1}, Guozheng Yang\textsuperscript{1}, Dongzhen Qi\textsuperscript{1}}
\IEEEauthorblockA{\textsuperscript{1}\textit{College of Electronic Engineering, National University of Defence Technology}, Hefei 230037, China}
\IEEEauthorblockA{\textsuperscript{2}\textit{Institute of Computing Technology, Chinese Academy of Sciences}, China}
\IEEEauthorblockA{\textsuperscript{3}\textit{University of Chinese Academy of Sciences}, China}
\IEEEauthorblockA{\textsuperscript{4}\textit{College of Computer Science and Technology, National University of Defence Technology}, Changsha 410000, China}
\IEEEauthorblockA{Corresponding author: yangguozheng17nudt.edu.cn}
}\maketitle

\begin{abstract}
Encrypted traffic classification is commonly evaluated under a range of experimental protocols, with classification performance reported as the primary measure.
However, the reported performance may be influenced by shortcuts left uncontrolled in the evaluation protocol, such as shortcut features and cross-partition sharing, rather than reflect generalizable traffic patterns.
And test-only feature transformations may further introduce train--test mismatch.
In this paper, we study how current evaluation protocols are affected by shortcut-related factors, including field-level shortcuts, cross-partition sharing, and train--test processing.
We compare feature importance, retrained feature removal, and field-only split sensitivity.
We also vary Source-IP overlap and packets per flow, compare test-only IP transformations with matched train--test transformations.
Experiments on three public datasets show that the performance of existing methods are affected by both cross-partition sharing and the consistency of train--test transformations.
Specifically, increasing Source-IP overlap from 0\% to 100\% raises accuracy by about 32--33 points before identifier masking, compared with only about 8 points after masking.
For two state-of-the-art pretrained models (i.e., YaTC and NetMamba), the 11--13 points loss caused by test-only IP zeroing is almost fully recovered when the same transformation is applied during both training and testing.

\end{abstract}

\section{Introduction}
\label{sec:introduction}

Encrypted traffic classification (ETC) supports network management and security monitoring.
Existing methods range from lightweight classifiers built on packet features~\cite{article,roy2022fast,xu2023fasttraffic} to pre-trained models such as ET-BERT, YaTC,  NetMamba~\cite{lin2022bert,zhao2023yet,wang2024netmamba}.
Recent methods report very high accuracy and F1 scores on commonly used benchmarks.
As benchmark performance becomes increasingly high, it becomes equally important to understand what contributes to these scores. In other words: if the factors the model relies on are not close to reality, then the performance of the model is actually artificially high.
This raises a question: What do current ETC evaluations actually measure?

In fact, the dataset split does more than divide samples.
It also determines which identifiers and repeated observations remain shared across the two partitions.
Identifier fields and repeated observations may form stable associations with labels, allowing a model to exploit shortcuts rather than learning broader traffic patterns~\cite{geirhos2020shortcut,zhao2025sweetdanger,arp2022and}.
Field masking and entity-based grouping address different parts of this problem: masking changes the information visible to the classifier, whereas grouping changes what can be shared across training and test.
Descriptions such as packet-level random split or flow-level split therefore do not fully specify the resulting evaluation condition.

A second issue appears when feature dependence is inferred from test-time perturbations.
If an IP field is removed or remapped only at test time, the input representation no longer matches the one used for training.
The resulting performance loss can therefore reflect both information removal and train--test mismatch.
Training and testing on the same transformed representation asks a different question: how well can the model perform when the transformation is applied consistently?
Distinguishing these two settings is necessary before a test-only drop is interpreted as evidence of feature dependence.

\begin{figure}[t]
\centering
\includegraphics[width=\columnwidth]{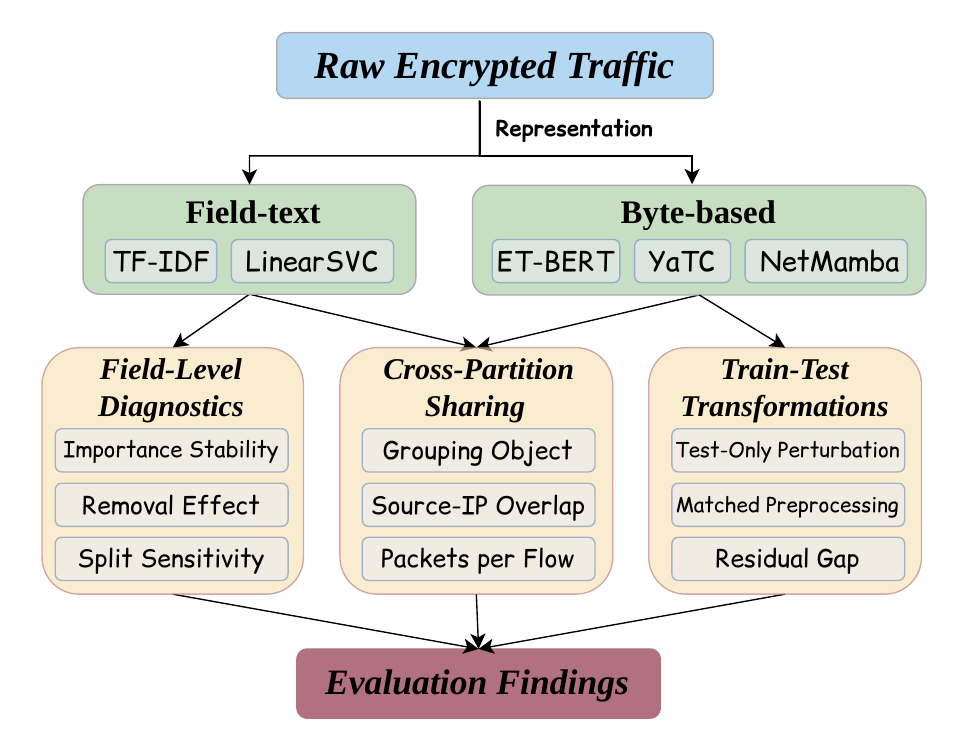}
\caption{Overview of the evaluation framework.}
\label{fig:framework}
\end{figure}

To separate these factors, we examine ETC evaluation from three complementary directions, as summarized in Fig.~\ref{fig:framework}:
(1) \textbf{Field-level diagnostics:} We compare importance stability, retrained removal effects, and field-only split sensitivity to characterize different forms of association between individual traffic fields and classification outcomes.
(2) \textbf{Cross-partition sharing:} We examine the grouping object and directly control source-IP overlap and the number of packets retained per flow to measure how shared identities and related observations affect evaluation results.
(3) \textbf{Train--test transformations:} We compare test-only IP perturbations with matched train--test preprocessing and measure the residual performance gap after matched training.

Experiments on three public datasets produce three main observations:
(1) Feature importance and field sensitivity are not equivalent. A field that appears unimportant in one setting can still form a strong split-sensitive label association in another;
(2) Increasing Source-IP overlap from 0\% to 100\% raises accuracy by about 32 points before identifier masking, but only about 8 points after masking;(3) Test-only IP zeroing causes an 11--13 points loss for YaTC and NetMamba, yet this loss is almost fully recovered under matched train--test processing, while ET-BERT/CSTNET retains a 47.1 points residual.
Together, these results show that ETC scores depend on both what is shared across partitions and how input transformations are applied.

The main contributions are as follows:
\begin{enumerate}[leftmargin=*,noitemsep,topsep=2pt]
\item We show that feature importance, retrained removal effect and field-only split sensitivity capture different properties of traffic fields.

\item We treat cross-partition sharing as an experimental variable by controlling Source-IP overlap and the number of packets retained per flow.
The results show that shared identities and repeated observations can substantially change measured performance.

\item We use matched train--test transformations to distinguish test-only perturbation sensitivity from performance under consistent preprocessing.
This comparison shows when a large test-only loss can be recovered through matched training and testing instead of being attributed directly to feature dependence.

\end{enumerate}

\section{Related Work}
\label{sec:related_work}
\subsection{Encrypted Traffic Classification}

Encrypted traffic classification has evolved from methods based on handcrafted packet and flow statistics~\cite{rezaei2019deep,article,roy2022fast,xu2023fasttraffic} to models that learn representations directly from traffic data.
ET-BERT learns contextualized representations from packet-level datagrams~\cite{lin2022bert}, while YaTC and NetMamba model information across multiple packets within a flow~\cite{zhao2023yet,wang2024netmamba}.
These models operate on representations ranging from individual packet bytes to multi-packet flow inputs.

Recent studies have questioned whether high benchmark performance translates to reliable classification beyond the evaluated train--test split.
The SoK by Wickramasinghe et al.~\cite{wickramasinghe2025sok} identifies dataset and evaluation design choices that can substantially affect encrypted traffic classification results.
Zhao et al.~\cite{zhao2025sweetdanger} further show that, under conventional evaluation settings, representation learning models can exploit correlations introduced during dataset construction and partitioning, leading to overly optimistic performance estimates.
These findings motivate closer examination of how ETC benchmarks are constructed, partitioned, and evaluated.

\subsection{Cross-Partition Sharing and Shortcut Learning}

Train--test partitioning determines which traffic entities and related observations may remain shared across the two partitions.
General studies of security machine learning have emphasized the need to separate related objects across training and test sets~\cite{arp2022and,pendlebury2019tesseract}, while recent ETC studies show that packet-level random splitting can place packets or related observations from the same flow on both sides of the partition~\cite{zhao2025sweetdanger}.
Such cases are often discussed as forms of train--test leakage, but they differ in the entities or observations that remain shared across partitions.
Flow- or group-aware splitting reduces specific forms of cross-partition sharing, but the resulting evaluation still depends on the grouping object used to construct the split.

Recent assessments of network-security and encrypted-traffic benchmarks also show that dataset construction and evaluation choices can strongly affect reported performance~\cite{flood2024bad,wickramasinghe2025sok}.
Shortcut learning occurs when a model relies on predictive correlations that do not capture the intended task semantics~\cite{geirhos2020shortcut}.
In encrypted traffic classification, such correlations can arise from protocol and header fields whose values are specific to a dataset, host, session, or collection environment.
For example, BiasSeeker~\cite{wang2026biasseeker} uses statistical association and field-level interventions to identify and validate potential shortcut features across traffic datasets.
Traffic filtering studies have also shown that preprocessing choices can change the information available to a classifier~\cite{lin2026filtering}.
Existing field-level diagnostics use different forms of evidence, including feature importance and performance changes after field removal or transformation.

However, these measurements characterize different aspects of field dependence and need not lead to the same conclusion.
Field intervention introduces a separate issue: when a transformation is applied only at test time, the model is evaluated on an input representation different from the one used during training.
The resulting performance drop may therefore reflect dependence on information carried by the modified field, sensitivity to the resulting train--test mismatch, or both.

\section{Methodology}
\label{sec:methodology}

\subsection{Problem Setup and Input Representations}
\label{sec:problem_setup}

We study encrypted traffic classification under different input representations and train--test partitioning rules.
Given a labeled traffic dataset, an evaluation protocol partitions it into training and test sets and measures classification performance on the held-out samples.
Our goal is to examine how this measured performance changes with the information available to the classifier, the entities shared across partitions, and the way input transformations are applied during training and testing.

We use two input representations with different measurement roles.
The \emph{field-text view} represents explicit packet fields as textual field--value tokens and encodes them with TF-IDF features.
It is used for field importance, feature removal, masking, and field-only diagnostics.
The \emph{byte-based view} uses packet byte sequences or multi-packet flow representations and is used to evaluate ET-BERT, YaTC, and NetMamba with their own model inputs.
The field-text analysis therefore diagnoses explicit field associations rather than serving as a surrogate explanation of the byte-based models.

The sample unit also varies with the evaluated representation.
Packet-level experiments treat individual packets as samples, whereas flow-level experiments construct one sample from multiple packets belonging to the same bidirectional parent flow.
For the latter, the main experiments use the first five packets of each flow.
The exact dataset, sample unit, and model assignment are reported in Section~\ref{sec:setup}.

All grouping keys are derived from the original traffic records before any masking or byte transformation is applied.
The same sample identities are then reused by the corresponding transformed conditions.
Thus, changing an input representation does not change how the underlying samples are assigned to training and test sets.

\subsection{Field-level Diagnostics}
\label{sec:field_diagnostics}

We use several complementary measurements to examine whether a traffic field forms a stable predictive association.
The first measurement is feature-importance stability over 22 packet fields.
Four importance methods are used: mutual information, coefficients from $L_1$-regularized logistic regression, random-forest feature importance, and single-feature decision-tree accuracy.
Each method is evaluated over three random seeds, and the five highest-ranked fields are recorded for each run.

For a field $f$, we define its stability as

\begin{equation}
\label{eq:stability}
\mathrm{Stability}(f)
=
\frac{1}{MR}
\sum_{m=1}^{M}
\sum_{r=1}^{R}
\mathbf{1}
\left[
f \in \mathrm{Top}\text{-}K(m,r)
\right],
\end{equation}

where $M=4$ is the number of importance methods, $R=3$ is the number of random seeds, and $K=5$.
A higher value indicates that the field is repeatedly ranked among the most predictive fields across methods and seeds.

Feature importance alone does not measure the performance contribution of a field.
We therefore retrain the classifier after removing each field.
Let $F_S(\mathrm{all})$ denote the Macro-F1 under split rule $S$ using all fields, and let $F_S(\mathrm{mask}\ f)$ denote the score after field $f$ is removed and the classifier is retrained.
The retrained removal effect is

\begin{equation}
\label{eq:removal}
U_f^{S}
=
F_S(\mathrm{all})
-
F_S(\mathrm{mask}\ f).
\end{equation}

We further measure whether a field alone remains predictive when the partitioning rule changes.
Let $A_f^{S}$ denote the Macro-F1 of a classifier trained using only field $f$ under split rule $S$.
Using Packet-Random as the reference partition, the field-only split sensitivity is

\begin{equation}
\label{eq:split_sensitivity}
D_{A,f}^{S}
=
A_f^{\mathrm{PR}}
-
A_f^{S},
\end{equation}

where $\mathrm{PR}$ denotes Packet-Random.
A larger $D_{A,f}^{S}$ indicates that the predictive value of field $f$ decreases more strongly when the corresponding grouping constraint is introduced.

As an auxiliary statistic, we also use

\begin{equation}
\label{eq:pf}
P_f = O_f Q_f,
\end{equation}

where $O_f$ describes the cross-partition overlap of field values and $Q_f$ describes their label specificity.
$P_f$ is used only to summarize whether a field provides stable value--label correspondences in a given dataset.

We define four cumulative masking levels:
\begin{description}
\setlength{\itemsep}{1pt}
\setlength{\parsep}{0pt}
\setlength{\topsep}{2pt}
\item[\textbf{L0}] Retains all original fields.
\item[\textbf{L1}] Masks MAC addresses and \texttt{frame.number}.
\item[\textbf{L2}] Additionally masks source and destination IP addresses, including IPv6 addresses.
\item[\textbf{L3}] Further masks transport-layer ports.
\end{description}

Each condition is trained independently on its corresponding masked representation.
The L0--L3 comparison therefore measures performance after masking and retraining rather than a test-time perturbation.

\subsection{Train--Test Partitioning and Cross--Partition Sharing}
\label{sec:partitioning}

Train--test partitioning determines which traffic entities can remain shared between the two sides of an evaluation.
We therefore describe each split by its grouping object rather than treating the split rules as an ordered hierarchy.

\emph{\textbf{Packet-Random}} assigns individual packets randomly to the two partitions.

\emph{\textbf{IP-pair grouped}} keeps packets with the same undirected source--destination IP pair on the same side.

\emph{\textbf{Source-IP grouped}} groups packets by \texttt{ip.src}.

\emph{\textbf{Parent-flow grouped}} keeps all packets from the same bidirectional parent flow together.
For flow-level samples, \emph{Flow-Random} randomly assigns the first-$k$ representation of each parent flow, whereas \emph{Initiator-disjoint} keeps flows with the same initiator on the same side while preserving parent-flow integrity.

These rules remove different forms of cross-partition sharing.
For each split, we therefore record the overlap of relevant entities, including Source IPs, IP pairs, and parent flows, instead of inferring overlap from the split name alone.
For a group-aware split, all samples belonging to the specified grouping object are assigned to one partition, so the corresponding training and test groups are disjoint.

We further vary cross-partition sharing directly.
In the Source-IP overlap experiment, the proportion of source identities appearing in both partitions is controlled from no overlap to complete overlap while keeping the remaining evaluation procedure fixed.
In the packets-per-flow experiment, the number of packets contributed by each parent flow is varied from one packet to multiple packets and finally to all available packets.
This controls how many related observations from the same flow can be distributed across a packet-level random split.

\subsection{Train--Test Transformations}
\label{sec:transformations}

Field perturbations can be evaluated either only at test time or consistently during both training and testing.
To distinguish these cases, we construct a train--test transformation matrix over four byte-based input versions.

\textbf{R} retains the original IP addresses, \textbf{P} applies per-flow IP and \textbf{Z} zeros the IP fields.
For $t\in\{P,Z\}$, we compare the original condition $R\!\rightarrow\!R$, the test-only condition $R\!\rightarrow\!t$, and the matched condition $t\!\rightarrow\!t$. We refer to their differences as the test-only drop, recovery after matched training and residual gap.

The byte transformations are distinct from the field-text masking levels defined in Section~\ref{sec:field_diagnostics}.
In particular, Z modifies only IP fields in the byte representation and is not equivalent to L2, while P removes stable cross-flow address identity rather than applying one global anonymization map.

Each training condition is optimized independently under the same model-specific training protocol.
The comparison between $R\!\rightarrow\!t$ and $t\!\rightarrow\!t$ therefore separates sensitivity to a test-only perturbation from the performance achievable after matched train--test processing.
\section{Experiments and Findings}
\label{sec:experiments}

We evaluate the factors defined in Section~\ref{sec:methodology} from three perspectives: field masking, cross-partition sharing, and train--test transformations.
We first establish how masking and grouping interact, and then use controlled experiments to examine the mechanisms behind the observed differences.

\subsection{Experimental Setup}
\label{sec:setup}

\paragraph{\textbf{Datasets and input representations}}
We conduct experiments on CSTNET-TLS~1.3, ISCXVPN~\cite{draper2016characterization}, and USTC-TFC~\cite{wang2017malware}.
The field-text view uses packet-level field--value tokens represented by 10,000-dimensional TF-IDF features with 1--3-grams.
The byte-based experiments use packet-byte inputs for ET-BERT and first-5-packet flow representations for YaTC and NetMamba, as summarized in Table~\ref{tab:setup}.
LinearSVC is used with the field-text representation and as a lightweight reference in the controlled experiments.

\begin{table}[t]
\centering
\caption{Datasets and input settings.}
\label{tab:setup}

\footnotesize
\setlength{\tabcolsep}{15pt}
\renewcommand{\arraystretch}{1.08}

\begin{threeparttable}
\begin{tabular}{@{}ccc@{}}
\toprule
Dataset
& \shortstack{Field-text view}
& \shortstack{Byte-based view} \\
\midrule
CSTNET-TLS 1.3
& Packet / TF-IDF
& Packet / bytes \\

ISCXVPN
& Packet / TF-IDF
& Flow / first-5 \\

USTC-TFC
& Packet / TF-IDF
& Flow / first-5 \\
\bottomrule
\end{tabular}

\begin{tablenotes}[flushleft]
\scriptsize
\item Field-text uses a 10,000-dimensional TF-IDF representation with 1--3-grams.
\item Flow inputs contain the first five packets of each parent flow.
\end{tablenotes}

\end{threeparttable}
\end{table}

\paragraph{\textbf{Models and evaluation}}

We evaluate LinearSVC together with three representative ETC models: ET-BERT~\cite{lin2022bert}, YaTC~\cite{zhao2023yet}, and NetMamba~\cite{wang2024netmamba}.
Model-specific inputs follow their corresponding preprocessing pipelines, and paired conditions use the same training configuration and checkpoint rule.

Macro-F1 is the primary metric for the masking, field-diagnostic, and transformation experiments.
The controlled Source-IP overlap and packets-per-flow experiments report accuracy or accuracy differences.
Most field and split experiments use five random seeds, whereas the ET-BERT Source-IP overlap and train--test transformation experiments use three.
Figures report mean $\pm$ one standard deviation unless stated otherwise, and metric differences are reported in percentage points.

\subsection{Masking and Train--Test Splitting}
\label{sec:mask_split}

\begin{figure}[t]
    \centering
    \includegraphics[width=\columnwidth]{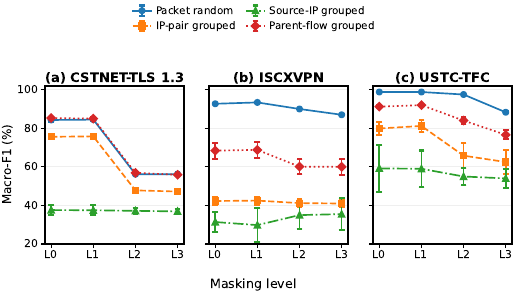}
    \caption{Interaction between field masking and train--test grouping.}
    \label{fig:mask_split}
\end{figure}

Figure~\ref{fig:mask_split} shows how cumulative field masking interacts with four train--test grouping rules across the three datasets.

CSTNET-TLS~1.3 shows the strongest interaction: Source-IP grouping reduces L0 Macro-F1 to $37.58\%$, after which additional masking changes performance only slightly.
On ISCXVPN, Parent-flow grouping instead decreases from $68.42\%$ at L0 to $60.08\%$ at L3.
USTC-TFC retains higher performance overall, but also shows substantial differences across grouping rules.
These results indicate that the effect of masking depends on the cross-partition sharing induced by the grouping rule.

\subsection{Field-level Diagnostics}
\label{sec:field_results}

Table~\ref{tab:field_diagnostics} compares importance stability, field-only split sensitivity, and retrained removal effects for eight representative fields.

The three diagnostics are not interchangeable because they characterize different properties of a field.
Importance stability measures how consistently a field is ranked as predictive, whereas the retrained removal effect measures its marginal contribution when the classifier is allowed to adapt to its absence.
Field-only split sensitivity instead measures how strongly the predictive value of a field depends on the partitioning rule.

\begin{table}[t]
\centering
\caption{Representative field-level diagnostic results.}
\label{tab:field_diagnostics}
\footnotesize
\setlength{\tabcolsep}{2.6pt}

\begin{threeparttable}
\begin{tabular*}{\columnwidth}{@{\extracolsep{\fill}}lcccccc@{}}
\toprule
\multirow{2}{*}{Field}
  & \multicolumn{2}{c}{CSTNET}
  & \multicolumn{2}{c}{ISCXVPN}
  & \multicolumn{2}{c}{USTC} \\
\cmidrule(lr){2-3}
\cmidrule(lr){4-5}
\cmidrule(lr){6-7}
  & Stab. & Rem. & Sens. & Rem. & Sens. & Rem. \\
\midrule
\texttt{ip.src}
  & 91.7 & 22.4
  & 13.7 & 1.2
  & 25.2 & 0.2 \\

\texttt{ip.dst}
  & 66.7 & -0.4
  & 12.4 & 1.0
  & 29.3 & 0.2 \\

\texttt{tcp.srcport}
  & 0.0 & 0.1
  & 43.1 & 0.0
  & 22.3 & 1.3 \\

\texttt{tcp.dstport}
  & 0.0 & 0.0
  & 48.0 & 0.1
  & 28.1 & 1.9 \\

\texttt{tcp.window\_size}
  & 50.0 & 2.7
  & 29.3 & 0.1
  & 7.4 & 0.7 \\

\texttt{tcp.seq}
  & 0.0 & 0.0
  & 21.6 & 0.2
  & 9.5 & 0.0 \\

\texttt{tcp.ack}
  & 0.0 & 0.0
  & 28.4 & 0.0
  & 10.4 & 0.1 \\

\texttt{frame.len}
  & 25.0 & 0.1
  & 10.8 & 0.0
  & 6.8 & 0.0 \\
\bottomrule
\end{tabular*}

\begin{tablenotes}[flushleft]
\scriptsize
\item Stab.: importance stability (\%);
\item Sens.: field-only split sensitivity, measured from Packet-Random to Parent-flow grouping;
\item Rem.: retrained removal effect.
\end{tablenotes}
\end{threeparttable}
\end{table}

These distinctions are visible in Table~\ref{tab:field_diagnostics}.
On CSTNET-TLS~1.3, \texttt{ip.src} has both high importance stability (91.7\%) and a large removal effect (22.4 points), whereas \texttt{ip.dst} has relatively high stability (66.7\%) but almost no removal effect.
On ISCXVPN, \texttt{tcp.dstport} shows a 48.0 points split sensitivity but only a 0.1 points removal effect.
Thus, a field can exhibit a strong or split-sensitive predictive association without making a correspondingly large marginal contribution to the retrained full-field classifier.

\subsection{Controlled Cross-Partition Sharing}
\label{sec:sharing}

\begin{figure}[t]
    \centering
    \includegraphics[width=\columnwidth]{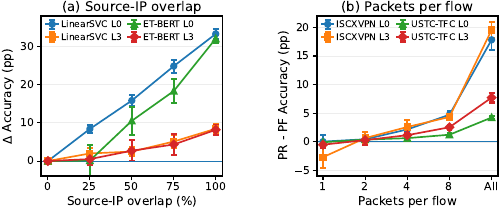}
    \caption{Controlled cross-partition sharing experiments.
    (a) Effect of Source-IP overlap.
    (b) Effect of packets per flow.}
    \label{fig:controlled_sharing}
\end{figure}

We next control two forms of cross-partition sharing directly.
Figure~\ref{fig:controlled_sharing} varies Source-IP overlap and the number of packets contributed by each parent flow.

\paragraph{\textbf{Source-IP overlap}}
At 100\% overlap, the L0 accuracy gain reaches $33.29$ points for LinearSVC and $32.05$ points for ET-BERT, whereas the corresponding L3 gains are only about $8$ points.
The effect of Source-IP sharing is therefore substantially reduced after identifier masking.

\paragraph{\textbf{Packets per flow}}
The Packet-Random--Parent-flow gap is near zero with one packet per flow and grows as additional packets from the same flow are retained.
At L0, the gap reaches $17.83$ points on ISCXVPN and $4.26$ points on USTC-TFC when all packets are used.

\subsection{Modern Models under Entity-Disjoint Splits}
\label{sec:modern_models}

Table~\ref{tab:modern_models} compares Flow-Random and initiator-disjoint evaluation using each model's corresponding input setting.
For each seed, the same common class set is used across models, and predictions outside this set are counted as errors.
Initiator-disjoint evaluation reduces Macro-F1 by $31.4$--$55.1$ points on ISCXVPN and by about $23$--$24$ points on USTC-TFC across the three models.
All paired 95\% confidence intervals remain above zero.

\begin{table}[t]
\centering
\caption{Macro-F1 under Flow-Random and initiator-disjoint}
\label{tab:modern_models}
\scriptsize
\setlength{\tabcolsep}{8pt}
\renewcommand{\arraystretch}{1.08}

\begin{threeparttable}
\begin{tabular}{@{}lccc@{}}
\toprule
Model
& \shortstack{Flow-Random}
& \shortstack{Initiator-disjoint}
& \shortstack{$\Delta$F1} \\
\midrule

\multicolumn{4}{c}{\textit{ISCXVPN}} \\
LinearSVC & $50.2\pm1.3$ & $18.9\pm1.9$ & $31.4$ \\
YaTC      & $78.3\pm0.8$ & $23.2\pm2.5$ & $55.1$ \\
NetMamba  & $72.6\pm3.6$ & $23.6\pm2.6$ & $49.0$ \\

\midrule

\multicolumn{4}{c}{\textit{USTC-TFC}} \\
LinearSVC & $97.8\pm0.3$ & $74.1\pm2.5$ & $23.7$ \\
YaTC      & $99.1\pm0.1$ & $75.2\pm2.5$  & $23.8$ \\
NetMamba  & $98.7\pm0.2$ & $75.4\pm2.9$  & $23.3$ \\

\bottomrule
\end{tabular}

\begin{tablenotes}[flushleft]
\scriptsize
\item $\Delta$F1 denotes Flow-Random minus initiator-disjoint Macro-F1.
\end{tablenotes}
\end{threeparttable}
\end{table}

\subsection{Test-only vs Matched Transformations}
\label{sec:transformation_results}

Table~\ref{tab:transformations} compares the original condition with test-only and matched versions of per-flow remapping and IP zeroing.

\begin{table}[t]
\centering
\caption{Effect of test-only and matched IP transformations on Macro-F1.}
\label{tab:transformations}

\footnotesize
\setlength{\tabcolsep}{5.2pt}
\renewcommand{\arraystretch}{1.12}

\begin{threeparttable}
\begin{tabular}{@{}cccccc@{}}
\toprule
\multirow{2}{*}{Model}
& \multirow{2}{*}{Transformation}
& \multicolumn{3}{c}{Macro-F1 (\%)}
& \multirow{2}{*}{\shortstack{Residual\\gap}} \\
\cmidrule(lr){3-5}
&
& Original
& Test-only
& Matched
& \\

\midrule

\multirow{2}{*}{ET-BERT\tnote{a}}
& Remap ($P$)
& \multirow{2}{*}{82.4}
& 5.4
& 33.0
& 49.4 \\
& Zero ($Z$)
&
& 4.3
& 35.3
& 47.1 \\

\midrule

\multirow{2}{*}{YaTC\tnote{b}}
& Remap ($P$)
& \multirow{2}{*}{99.4}
& 89.0
& \textbf{99.4}
& \textbf{0.0} \\
& Zero ($Z$)
&
& 88.4
& \textbf{99.5}
& \textbf{-0.1} \\

\midrule

\multirow{2}{*}{NetMamba\tnote{b}}
& Remap ($P$)
& \multirow{2}{*}{99.4}
& 88.2
& \textbf{99.2}
& \textbf{0.2} \\
& Zero ($Z$)
&
& 86.4
& \textbf{99.4}
& \textbf{0.0} \\

\bottomrule
\end{tabular}

\begin{tablenotes}[flushleft]
\scriptsize
\item[a] CSTNET-TLS 1.3, packet-level input.
\item[b] USTC-TFC, first-5-packet flow input.
\end{tablenotes}

\end{threeparttable}
\end{table}

On USTC-TFC, YaTC and NetMamba lose 11--13 Macro-F1 points under test-only IP zeroing, but matched training restores performance to approximately the original level.
Per-flow remapping shows the same pattern. For ET-BERT on CSTNET-TLS1.3, matched zeroing recovers 31.0 points but leaves a 47.1-point gap relative to $R\!\rightarrow\!R$. Thus, the loss remaining after retraining can differ substantially from the test-only drop.
\section{Discussion}
\label{sec:discussion}

\subsection{Test-Only Perturbation vs. Matched Preprocessing}

Test-only and matched transformations answer different evaluation questions.
A test-only transformation measures the sensitivity of a fixed trained model to a change in its input representation.
Matched preprocessing instead measures the performance achievable when training and testing use the same transformed representation.

This distinction changes how perturbation results should be interpreted.
For YaTC and NetMamba, most of the loss caused by test-only IP transformations disappears after matched training, indicating that the test-only drop largely reflects sensitivity to representation mismatch rather than an irrecoverable loss of predictive information.
ET-BERT retains a substantially larger residual after matched processing.
This residual may arise from information removed by the transformation, incompatibility with the pretrained representation, or both

\indent\textbf{\textit{Takeaway 1:}}
\textit{A test-only performance drop should not by itself be interpreted as the amount of information contributed by the modified field.}

\subsection{Implications for ETC Evaluation}

Our results suggest that an ETC evaluation should specify more than the classifier, dataset, and nominal train--test split.
At minimum, the evaluation protocol should report the sample unit and the grouping object used to construct the split, because different choices preserve different relationships between training and test samples.
Where relevant, the resulting cross-partition overlap of entities such as hosts, IP pairs, flows or initiators should also be reported rather than inferred from the split name alone.

Feature interventions require a similar distinction.
When a field is modified only at test time, the observed performance change reflects the response to a test representation different from that used during training.
Matched preprocessing provides a complementary measurement by applying the same transformation during training and testing.

\indent\textbf{\textit{Takeaway 2:}}
\textit{ETC evaluations should report relevant cross-partition sharing and distinguish test-only interventions from matched train--test preprocessing.}

\section{Conclusion}
\label{sec:conclusion}

This work examines what encrypted traffic classification benchmarks measure when shortcut features, cross-partition sharing, and train--test preprocessing are varied.
Field-level diagnostics show that feature importance and retrained removal need not assign the same role to a field, while field-only experiments show that predictive value can change sharply with the partitioning rule.
Controlled Source-IP overlap and packets-per-flow experiments further show that shared identities and repeated observations can substantially change measured performance.

The train--test transformation experiments reveal a separate source of ambiguity.
For YaTC and NetMamba on USTC-TFC, matched training recovers nearly all of the 11--13 points Macro-F1 drop caused by test-only IP zeroing. A test-only drop therefore does not measure the performance loss that remains after retraining.
ET-BERT retains a substantial residual under matched processing, although this residual may combine information removal with pretraining distribution shift.

Reliable ETC evaluation therefore requires more than reporting a classifier and a train--test split.
It should specify the sample unit, the grouping object, what remains shared across partitions, and whether feature transformations are applied only at test time or consistently during both training and testing.

\bibliographystyle{IEEEtran}
\bibliography{references}

\end{document}